\documentclass[onecolumn, 12pt, draftcls]{IEEEtran}
\usepackage{amsmath,amssymb,amsfonts}
\def \mbf {\mathbf}
\renewcommand{\IEEEQED}{\IEEEQEDopen}
\newtheorem{theorem}{Theorem}
\newtheorem{lemma}{Lemma}
\newtheorem{prop}{Proposition}

\newtheorem{ex}{Example}
\newtheorem{con}{Conjecture}

\usepackage{graphicx}
\ifCLASSINFOpdf
\else
\fi
\begin{document}
%
\title{Efficient decoding algorithm using triangularity of $\mbf{R}$ matrix of QR-decomposition}
%
%
%

\author{In~Sook~Park

\thanks{The author is with the BK Institute of Information and Technology, Division of Electrical Engineering, Department
of Electrical Engineering and Computer Science, KAIST, Daejeon,
Korea [e-mail: ispark@amath.kaist.ac.kr; ispark@kaist.ac.kr].}}
\maketitle

\begin{abstract}
An efficient decoding algorithm named `divided decoder' is proposed in this paper. Divided decoding can be combined with any decoder using QR-decomposition and offers different pairs of performance and complexity. Divided decoding provides various combinations of two or more different searching algorithms. Hence it makes flexibility in error rate and complexity for the algorithms using it. We calculate  diversity orders and upper bounds of error rates for typical models when these models are solved by divided decodings with sphere decoder, and discuss about the effects of divided decoding on complexity. Simulation results of divided decodings combined with a sphere decoder according to different splitting indices correspond to the theoretical analysis.

\end{abstract}


%
\IEEEpeerreviewmaketitle

\section{Introduction}
%

\IEEEPARstart{T}{o} obtain high data rate and spectral efficiency, communication systems require a detector the error rate of which is as close to that of the maximum likelihood (ML) solution as possible with a tolerable complexity. In most cases the additive noise vector is assumed to be Gaussian with mean zero-vector and detecting original signal from a received signal turns into solving an integer least-squares problem. This paper proposes a method solving the integer least-squares problem which is finding $\hat{\mbf{s}}$ such that \begin{equation}\label{e1}\hat{\mbf{s}}=\min_{\mbf{s}\in D}\|\mbf{x}-\mbf{Hs}\|^2\end{equation} where $D$ is a set of $n$-dimensional complex vectors whose real and imaginary parts are integers (or discrete numbers), $\mbf{x}$ is an $m$-dimensional complex vector, and $\mbf{H}$ is an $m\times n$ complex matrix. The exact solution of (\ref{e1}) is ML solution when $\mbf{x}-\mbf{Hs}$ is an $m \times 1$ Gaussian random vector whose mean is $\mbf{0}_{m\times 1}$. The brute-force search visits all the points of $D$, which makes the complexity grow exponentially in $n$. Sphere decoding (SD) \cite{Vi1, Vi2, Ag, Ho, Da}, a depth first tree search within a sphere which can shrink with each new candidate during search process, is known to find the exact solution of (\ref{e1}) but reduce considerably the complexity so that it finds very often the solution within real time when the brute-force search can not. The efficient search
strategies \cite{Po, Fi, Sc} are employed by both real and complex sphere decoders \cite{Ri}. Usually before starting search process SD calculates the initial radius but, as noted in \cite{Ri}, when Schnorr-Euchner \cite{Sc} strategy is used the radius of the Babai point \cite{Ag} is enough for good start of search and the time required for the initial radius estimation is saved. The expected complexity of SD is known to be approximately polynomial for a wide range of signal-to-noise ratios (SNRs) and numbers of antennas \cite{Ha1, Ha2}. But it still depends on SNR and has more portion of high ordered terms in the dimension of the vector in search.

Algorithms finding near ML solutions with the advantage of complexity reduction have been suggested for recent decades. Among them, the M-algorithm combined with QR-decomposition (QRD-M) (\cite{Ki, Yu}) has performance almost the same as ML when the value of M is not less than the constellation number. For fixed M, the computation amount of QRD-M is independent of SNR and the condition number of channel matrices, and is polynomial in the dimension of the vector to be searched. But, for almost the same performance the expected computation amount of SD is much less than that of QRD-M though the maximum computation amount of SD is more than two times of the maximum computation amount of QRD-M \cite{Dai}. Detection with the aid of lattice reduction (LR) is another approach: LR helps SD to reduce the complexity \cite{Ag} when the channel matrix is ill-conditioned and aids linear detection or successive interference cancelation (SIC) to have better performances \cite{Ya, Wi, Wu2}. Though, checking the validity of every searched point adds computational load and calculating Log-likelihood ratio (LLR) is still burdensome for the LR aided detections. Fixed-complexity sphere decoder \cite{BBa} (FSD) is SD within a subset of the domain to be searched and visits only a fixed number of lattice points. FSD with a proper restricted domain has a near ML performance with a fixed complexity for each set of $m, n$ and constellation.

Nulling and cancelling with optimal orderings, i.e. zero-forcing with ordered successive interference cancellation (ZF-OSIC) and minimum mean square error with ordered successive interference cancellation (MMSE-OSIC), \cite{Fos} are sorts of standards and give bases for developing advanced decoding algorithms. ZF-OSIC and MMSE-OSIC both are performed efficiently and have computation amount reduced by employing QR-decomposition (QRD) or sorted QRD (SQRD) \cite{Wu1}. Nulling and cancellings and near ML algorithms above perform QRD before searching process. (Instead of QRD Cholesky decomposition is frequently used.) In practice, ZF-OSIC and MMSE-OSIC are available in error rate sense for higher modulations than QPSK when the number of transmit streams is no more than 4. If the number of transmit streams is more than 4 with high modulation, decoding algorithms performing in real time with lower error rate than nulling and cancellings are required. To support this requirement, we propose a simple method called `divided decoding' which utilizes the properties of the resultant matrices of QRD (or Cholesky decomposition) and combines with any given searching algorithms. Divided decoding can provide various modifications or combinations of searching algorithms which are known or to be appeared.

The remainder is composed of five sections as follows.
In Section \ref{sec2} we describe a basic system model to solve. In Section \ref{sec3} we introduce the idea of divided decoding and the possible combination forms of the divided decoding and other algorithms. Section \ref{sec4} provides diversity orders and upper bounds of the error probabilities for some typical models by summing up pairwise error probabilities when the divided decoding is combined with SD, and a discussion of complexity reduction effects of the divided decoding. Section \ref{sec5} presents simulation results supporting the analyses in section \ref{sec4} by showing the way of transitions of bit error rate (BER) and complexity curves versus SNR according to the splitting index set, and compares divided decodings based on SD with Lenstra Lenstra and Lov\'{a}sz (LLL) LR \cite{Len} aided SIC's. In Section \ref{sec6} there is a conclusion.

\section{System Model} \label{sec2}
An original signal vector $\mbf{s}$ belong to $D$, a finite subset of an $n$ dimensional lattice, passes through a channel and is measured as an $m$ dimensional vector $\mbf{x}$, then the relation of $\mbf{s}$ and $\mbf{x}$ is modeled by \begin{equation}\label{e2} \mbf{x}=\mbf{Hs}+\mbf{n}\end{equation} where $\mbf{H}$ is an $m\times n$ channel matrix whose distribution is arbitrary and the elements of $\mbf{n}$ are assumed to be independently identically distributed (i.i.d.) circularly symmetric complex normal variables with mean zero and variance $\sigma^2$. Usually, for $q-$QAM constellations $D$ is the Cartesian product of $n$ copies of $q$ lattice points. $(\ref{e1})$ is the ML solution of $(\ref{e2})$. $(\ref{e2})$ is transformed to a real system, if the decoding algorithm used is based on real number calculations.

To describe the algorithm we propose, we need the following notation: The sub matrix composed of the elements in rows $a$ through $b$ of columns $c$ through $d$ of a matrix $\mbf{A}$ is denoted by $\mbf{A}[a:b][c:d]$. When $\mbf{v}$ is a column vector, the sub-vector composed of the elements in rows $a$ through $b$ of $\mbf{v}$ is denoted by $\mbf{v}[a:b]$.

\section{Divided decoding} \label{sec3}
\subsection{The Idea of Divided decoding}
First, $\mbf{H}$ is decomposed into $\mbf{QR}$ by QRD where $\mbf{Q}$ is a $m\times n$ matrix of orthonormal columns which is the first $m\times n$ partial matrix of a unitary matrix and $\mbf{R}$ is an $n\times n$ upper-triangular matrix with non-negative diagonal entries. $\mbf{QR}$ is called the thin factorization of $\mbf{H}$. To improve the performance of the algorithm presented below, either the columns of $\mbf{H}$ are reordered in increasing order of the Euclidean norm before QRD or $\mbf{H}$ is decomposed by sorted QRD (SQRD) which is a QRD intervened by sorting process of columns. SQRD is found in \cite{Wu1}. SQRD is more effective for performance improvement and we use SQRD in the followings. We let $\mbf{y}=\mbf{Q}^{*}\mbf{x}$ and $\mbf{z}=\mbf{Q}^{*}\mbf{n}$ where $\mbf{Q}^{*}$ is the conjugate transpose of $\mbf{Q}$. Then (\ref{e2}) is reformulated as \begin{equation}\label{e3} \mbf{y}=\mbf{Rs}+\mbf{z}\end{equation} where $\mbf{z}$ is statistically equivalent to $\mbf{n}$ i.e. the elements of $\mbf{z}$ are i.i.d. circularly symmetric complex normal variables with mean zero and variance $\sigma^2$. For any $1\leq i,j\leq n$, the inner product of $i$th and $j$th columns of $\mbf{R}$ is equal to the inner product of $i$th and $j$th columns of $\mbf{H}$. Hence the SNR for each symbol of $\mbf{s}$ is unchanged.

The simplest version of divided decoding is as follows: i) For any $i_0 (1\leq i_0 < n)$, let (\ref{e3}) be split into \begin{equation}\label{e4} \mbf{y}_1=\mbf{R}_1\begin{bmatrix}\mbf{s}_1 \\ \mbf{s}_2 \end{bmatrix} + \mbf{z}_1, \ \ \mbf{y}_2=\mbf{R}_2\mbf{s}_2+\mbf{z}_2 \end{equation} where $\mbf{R}_1=\mbf{R}[1:i_0][1:n], \mbf{y}_1=\mbf{y}[1:i_0], \mbf{s}_1=\mbf{s}[1:i_0], \mbf{z}_1=\mbf{z}[1:i_0], \mbf{R}_2=\mbf{R}[i_0+1:n][i_0+1:n], \mbf{y}_2=\mbf{y}[i_0+1:n], \mbf{s}_2=\mbf{s}[i_0+1:n]$, and $\mbf{z}_2=\mbf{z}[i_0+1:n]$.
First, find $\mbf{s}_2$ minimizing $\|\mbf{y}_2-\mbf{R}_2\mbf{s}_2\|^2$ by applying one of SD, M-algorithm and other near ML algorithms. Let $\hat{\mbf{s}}_2$ denote this point and calculate $\tilde{\mbf{y}}_1=\mbf{y}_1-\mbf{R}_1[1:i_0][i_0+1:n]\hat{\mbf{s}}_2$. Secondly, find $\mbf{s}_1$, denoted by $\hat{\mbf{s}}_1$, minimizing $\|\tilde{\mbf{y}}_1-\mbf{R}_1[1:i_0][1:i_0]\mbf{s}_1\|^2$ by applying one of SD, M-algorithm and other near ML algorithms. $\begin{bmatrix}\hat{\mbf{s}}_1 \\ \hat{\mbf{s}}_2 \end{bmatrix}$ is an approximate solution of (\ref{e1}).

Method (i) is extended as follows: ii) (\ref{e3}) is split into more than two equations. Given $i_0, i_1, \dots, i_k$ $(1\leq i_0<i_1<\cdots<i_k<n)$, let $i_{-1}=0, i_{k+1}=n$ and then, for $1\leq f \leq k+2$, let $\mbf{R}_f=\mbf{R}[i_{f-2}+1:i_{f-1}][i_{f-2}+1:n], \mbf{y}_f=\mbf{y}[i_{f-2}+1:i_{f-1}], \mbf{s}_f=\mbf{s}[i_{f-2}+1:i_{f-1}], \mbf{z}_f=\mbf{z}[i_{f-2}+1:i_{f-1}]$. Then (\ref{e3}) is split into $k+2$ equations as follows: for $1\leq f \leq k+2$ \begin{equation}\label{e5} \mbf{y}_f=\mbf{R}_f\begin{bmatrix}\mbf{s}_f\\ \vdots \\ \mbf{s}_{k+2}\end{bmatrix}+\mbf{z}_f.\end{equation} We find $\mbf{s}_{k+2}$, denoted by $\hat{\mbf{s}}_{k+2}$, minimizing $\|\mbf{y}_{k+2}-\mbf{R}_{k+2}\mbf{s}_{k+2}\|^2$. Starting from $f=k+1$, compute $\tilde{\mbf{y}}_f=\mbf{y}_f-\mbf{R}_f[1:i_{f-1}-i_{f-2}][i_{f-1}-i_{f-2}+1:n-i_{f-2}]{\begin{bmatrix}\hat{\mbf{s}}_{f+1} & \cdots & \hat{\mbf{s}}_{k+2}\end{bmatrix}}^T$ and detect $\hat{\mbf{s}}_f$ minimizing $\|\tilde{\mbf{y}}_f-\mbf{R}_f[1:i_{f-1}-i_{f-2}][1:i_{f-1}-i_{f-2}]\mbf{s}_f\|^2$ repeatedly with decreasing $f$ one by one until $f=1$. Consequently, we obtain $\hat{\mbf{s}}_1, \dots, \hat{\mbf{s}}_{k+2}$. ${\begin{bmatrix}\hat{\mbf{s}}_1 & \cdots & \hat{\mbf{s}}_{k+2}\end{bmatrix}}^T$ is an approximate solution of (\ref{e1}).

If $i_0=1, i_1=2, \dots, i_{k=n-2}=n-1$ ($i_0=2, i_1=4, \dots, i_{k=n/2-2}=n-2$ if (\ref{e2}) is a real version of the original complex system) then the above method is the same as ZF-OSIC. As the number of split equations is increasing, the computation amount decreases but the error rate increases.

\subsection{Divided decoding with Quasi MMSE extension}
As described in \cite{Wu1}, the MMSE filter output $\tilde{\mbf{s}}_{MMSE}$ is reformulated by \begin{equation}\label{e6}\tilde{\mbf{s}}_{MMSE}=\big{(}\bar{\mbf{H}}^{*}\bar{\mbf{H}}\big{)}^{-1}\bar{\mbf{H}}^{*}\bar{\mbf{x}}\end{equation} where $\bar{\mbf{H}}$ and $\bar{\mbf{x}}$ are \begin{equation}\label{e7}\bar{\mbf{H}}=\begin{bmatrix}\mbf{H} \\ \sigma\mbf{I}_n\end{bmatrix} \mbox{ and }\bar{\mbf{x}}=\begin{bmatrix}\mbf{x} \\0_{n\times 1} \end{bmatrix}.\end{equation} We can reconstruct an extended system of (\ref{e2}) as follows: \begin{equation}\label{e8}\bar{\mbf{x}}=\bar{\mbf{H}}\mbf{s}+\bar{\mbf{n}}\end{equation} where $\bar{\mbf{n}}=\begin{bmatrix}\mbf{n} \\ -\sigma\mbf{s}\end{bmatrix}$ and $\bar{\mbf{n}}$ is assumed to be a Gaussian noise vector. We ignore $-\sigma\mbf{s}$ and regard it as a noise vector.

Instead of $\mbf{H}$, perform SQRD on $\bar{\mbf{H}}$ to obtain $\bar{\mbf{H}}=\bar{\mbf{Q}}\bar{\mbf{R}}$ and multiply (\ref{e8}) by $\bar{\mbf{Q}}^{*}$ to obtain \begin{equation}\label{e9}\bar{\mbf{y}}=\bar{\mbf{R}}\mbf{s}+\bar{\mbf{z}}\end{equation} where $\bar{\mbf{y}}=\bar{\mbf{Q}}^{*}\bar{\mbf{x}}$ and $\bar{\mbf{z}}=\bar{\mbf{Q}}^{*}\bar{\mbf{n}}$. If we search $\hat{\mbf{s}}_M=\min_{\mbf{s}\in D}\|\bar{\mbf{y}}-\bar{\mbf{R}}\mbf{s}\|^2$ by SD then $\hat{\mbf{s}}_M$ is a near ML solution which has almost negligible performance loss in comparison with ML solution.

Quasi MMSE extension is a generalization of MMSE extension as follows \cite{Pa}: \begin{equation}\label{e10}\bar{\mbf{x}}=\begin{bmatrix}\mbf{H} \\ \epsilon\sigma\mbf{I}_n\end{bmatrix}\mbf{s}+\begin{bmatrix}\mbf{n} \\ -\epsilon\sigma\mbf{s}\end{bmatrix}\end{equation} where $\epsilon$ is a positive real number. Let $\bar{\mbf{H}}_\epsilon=\begin{bmatrix}\mbf{H} \\ \epsilon\sigma\mbf{I}_n\end{bmatrix}$ and $\hat{\mbf{s}}_\epsilon=\min_{\mbf{s}\in D}\|\bar{\mbf{x}}-\bar{\mbf{H}}_\epsilon\mbf{s}\|^2$, then $\hat{\mbf{s}}_{\epsilon=1.0}=\hat{\mbf{s}}_M$. The performances of $\hat{\mbf{s}}_\epsilon$ for several $\epsilon$'s and the effects of Quasi MMSE extension on closest point search in complexity are described in \cite{Pa}. When $\epsilon=\frac1{\sqrt{2}}, \frac1{\sqrt{3}}$ the performance of $\hat{\mbf{s}}_\epsilon$ for low SNR range is better than $\hat{\mbf{s}}$ (ML solution) but the complexity required to find $\hat{\mbf{s}}_\epsilon$ by using SD is far lower than that to find $\hat{\mbf{s}}$.  This scenario is expected to be right for other $\epsilon$'s between 0 and 1.0. For $0\leq \epsilon \leq \sqrt{2}$ $\hat{\mbf{s}}_\epsilon$ has almost the same BER with $\hat{\mbf{s}}$, and as $\epsilon$ increases within at least $\sqrt{3}$ the computation amount decreases.

By SQRD on $\bar{\mbf{H}}_\epsilon$ we obtain $\bar{\mbf{H}}_\epsilon=\bar{\mbf{Q}}_\epsilon\bar{\mbf{R}}_\epsilon$ and get \begin{equation}\label{e11}\bar{\mbf{y}}_\epsilon=\bar{\mbf{R}}_\epsilon\mbf{s}+\bar{\mbf{z}}_\epsilon\end{equation} where $\bar{\mbf{y}}_\epsilon=\bar{\mbf{Q}}_\epsilon^{*}\bar{\mbf{x}}$ and $\bar{\mbf{z}}_\epsilon=\bar{\mbf{Q}}_\epsilon^{*}\begin{bmatrix}\mbf{n} \\ -\epsilon\sigma\mbf{s}\end{bmatrix}$. Although $\bar{\mbf{z}}_\epsilon$ contains the unknown signal $
\mbf{s}$, $\bar{\mbf{z}}_\epsilon$ is assumed to be a Gaussian noise vector with $\epsilon\sigma\mbf{s}$ ignored. $\hat{\mbf{s}}_\epsilon=\min_{\mbf{s}\in D}\|\bar{\mbf{y}}_\epsilon-\bar{\mbf{R}}_\epsilon\mbf{s}\|^2$ and $\hat{\mbf{s}}_\epsilon$ can be found by SD. (\ref{e11}) can be divided in the same way as (\ref{e5}) and approximate solutions to $\hat{\mbf{s}}_\epsilon$ can be obtained by searching all the sub-vectors.

\subsection{Hybrid Algorithms via Divided decoding}
Various combinations of two or more detection algorithms can be employed to find solutions after splitting equations (\ref{e3}), (\ref{e9}), (\ref{e11}) into the form of (\ref{e5}). For example, if starting from (\ref{e4}) firstly find $\hat{\mbf{s}}_2$ by SD and cancel $\hat{\mbf{s}}_2$ from $\mbf{y}_1$ by calculating $\tilde{\mbf{y}}_1=\mbf{y}_1-\mbf{R}_1[1:i_0][i_0+1:n]\hat{\mbf{s}}_2$. Then find $\hat{\mbf{s}}_1$ by SIC. Since SINR of $\mbf{s}_2$ is roughly no less than that of $\mbf{s}_1$ by column reordering, this hybrid algorithm reduces the error propagation against the pure SIC and reduces the complexity against SD. This combination is in fact the same with the case of finding each sub-vector solution by SD from (\ref{e5}) with $j_0=1, j_1=2, \dots, j_{i_0-2}=i_0-1, j_{i_0-1}=i_0$. Instead of SD and SIC, another combination like M-algorithm and SIC, SD and M-algorithm, or fixed-complexity SD and SIC can be applied.

\section{Error probability and Complexity} \label{sec4}
\subsection{Error probability}
It is well-known that MMSE-SIC or MMSE-OSIC, which is the original version and not the modified version of back substitution via transforming the channel matrix into a triangular one, can achieve the capacity of a given system \cite{Ts}. Back substitution after MMSE-SQRD or SQRD of $\mbf{H}$ and multiplying $\bar{\mbf{Q}}^{*}$ or $\mbf{Q}^{*}$ can not avoid some information loss due to ignoring strictly upper triangular part at each decision step and fails to achieve the capacity of the system. But the difference presented in BER curves of the former and the latter is small, because the degree of freedom at each step of decision which is related to the diversity order is an important factor of the error rate and the two have the same degree of freedom at each decision.

Divided decoding with nontrivial split can not achieve the capacity of a given system. Even in the case that the search algorithm for each sub-vector has ML performance, divided decoding with nontrivial split has information loss. The total achievable rate of method (ii) is \begin{equation}\label{e12}C_{d}=E_{\mbf{R}}\Big{[}\sum_{f=1}^{k+2}\log_2det(\mbf{I}_{n_f}+\frac1{\sigma^2}\tilde{\mbf{R}}_f\mbf{P}_f\tilde{\mbf{R}}_f^{*})\Big{]}\end{equation} where $\mbf{P}_f$ is the covariance matrix of $\mbf{s}_f$, $n_f=i_{f-1}-i_{f-2}$, and $\tilde{\mbf{R}}_f=\mbf{R}[i_{f-2}+1:i_{f-1}][i_{f-2}+1:i_{f-1}]$. There is information loss related to $\mbf{R}[1:i_{f-2}][i_{f-2}+1:i_{f-1}]$. Here $E_{\mbf{R}}[\cdot]$ denotes the expectation over $\mbf{R}$.

An upper bound of the error probability of a system can be obtained via the union bound of each pairwise probability, i.e. the average error rate of (\ref{e1}) is \begin{equation}P_{err}\leq E_{\mbf{s}\in D}\begin{bmatrix}\sum_{\mbf{s}^{'}\in D, \mbf{s}^{'}\neq\mbf{s}}P(\mbf{s}\rightarrow\mbf{s}^{'})\end{bmatrix}.\end{equation} $E_{\mbf{s}\in D}[\cdot]$ denotes the expectation over $\mbf{s}$ and $P(\mbf{s}\rightarrow\mbf{s}^{'})$ the probability that $\mbf{s}$ is mistaken for a different vector $\mbf{s}^{'}$. For each fixed (or estimated at the receiver) $\mbf{H}$ \begin{equation} P(\mbf{s}\rightarrow\mbf{s}^{'})=\frac1{\sqrt{2\pi}}\int_{\sqrt{\|\mbf{H}(\mbf{s}-\mbf{s}^{'})\|^2/(2\sigma^2)}}^\infty e^{-t^2/2}dt\end{equation} when we use a detector finding the ML solution. We let $\mathfrak{Q}(\alpha):=\frac1{\sqrt{2\pi}}\int_{\alpha}^\infty e^{-t^2/2}dt$. If we use the divided decoding which splitting (\ref{e3}) into the form (\ref{e5}) with $k\geq 0$ then for each sub-vector $\mbf{s}_f$ the pairwise probability $P(\mbf{s}_f\rightarrow \mbf{s}_f^{'})$ is calculated as follows: for  $f=k+2$, $P(\mbf{s}_{k+2}\rightarrow \mbf{s}_{k+2}^{'})=\mathfrak{Q}\begin{pmatrix}\sqrt{\frac{\|\tilde{\mbf{R}}_{k+2}(\mbf{s}_{k+2}-\mbf{s}_{k+2}^{'})\|^2}{2\sigma^2}}\end{pmatrix}$, and for $f<k+2$, \begin{equation*}\begin{split}P(\mbf{s}_f\rightarrow\mbf{s}_f^{'})&=P\big{(}\mbf{s}_f\rightarrow\mbf{s}_f^{'}\big{|}\hat{\mbf{s}}_{k+2}=\mbf{s}_{k+2}, \hat{\mbf{s}}_{k+1}=\mbf{s}_{k+1},\dots, \hat{\mbf{s}}_{f+1}=\mbf{s}_{f+1}\big{)}P(\hat{\mbf{s}}_{k+2}=\mbf{s}_{k+2}, \hat{\mbf{s}}_{k+1}=\\&\mbf{s}_{k+1},\dots, \hat{\mbf{s}}_{f+1}=\mbf{s}_{f+1})+P\begin{pmatrix}\mbf{s}_f\rightarrow\mbf{s}_f^{'}\Big{|}\{\hat{\mbf{s}}_{k+2}=\mbf{s}_{k+2}, \hat{\mbf{s}}_{k+1}=\mbf{s}_{k+1},\dots, \hat{\mbf{s}}_{f+1}=\mbf{s}_{f+1}\}^c\end{pmatrix}\\&\times P\begin{pmatrix}\{\hat{\mbf{s}}_{k+2}=\mbf{s}_{k+2}, \hat{\mbf{s}}_{k+1}=\mbf{s}_{k+1},\dots, \hat{\mbf{s}}_{f+1}=\mbf{s}_{f+1}\}^c\end{pmatrix}.\end{split}\end{equation*} We have \[P(\mbf{s}_f\rightarrow\mbf{s}_f^{'}|\hat{\mbf{s}}_{k+2}=\mbf{s}_{k+2}, \hat{\mbf{s}}_{k+1}=\mbf{s}_{k+1},\dots, \hat{\mbf{s}}_{f+1}=\mbf{s}_{f+1})=\mathfrak{Q}\begin{pmatrix}\sqrt{\frac{\|\tilde{\mbf{R}}_f(\mbf{s}_f-\mbf{s}_f^{'})\|^2}{2\sigma^2}}\end{pmatrix}\] and \[P\begin{pmatrix}\begin{pmatrix}\mbf{s}_f\rightarrow\mbf{s}_f^{'}\big{|}\{\hat{\mbf{s}}_{k+2}=\mbf{s}_{k+2}, \hat{\mbf{s}}_{k+1}=\mbf{s}_{k+1},\dots, \hat{\mbf{s}}_{f+1}=\mbf{s}_{f+1}\}^c\end{pmatrix}\end{pmatrix}\leq\mathfrak{Q}\begin{pmatrix}\sqrt{\frac{\|\tilde{\mbf{R}}_f(\mbf{s}_f-\mbf{s}_f^{'})\|^2}{2\sigma^2}}\end{pmatrix}\] because the middle of the distribution of $\tilde{\mbf{y}}_f$, which is the mean of $\tilde{\mbf{y}}_f$, under the condition that $\begin{bmatrix}\hat{\mbf{s}}_{f+1} && \cdots && \hat{\mbf{s}}_{k+2}\end{bmatrix}\neq\begin{bmatrix}\mbf{s}_{f+1} && \cdots && \mbf{s}_{k+2}\end{bmatrix}$ is not $\tilde{\mbf{R}}_f\mbf{s}_f$. Thus, we have the following inequality.
\begin{prop}
\label{prop1}
For fixed $\mbf{H}$, the error probability $P_{err}$ for the divided decoding (\ref{e5}) satisfies that \begin{equation}\label{e15}P_{err}\leq \sum_{f=1}^{k+2}E_{\{\mbf{s}_f\in D_f\}}\begin{bmatrix}\sum_{\{\mbf{s}_f^{'}\in D_f,\mbf{s}_f^{'}\neq\mbf{s}_f\}}\mathfrak{Q}\begin{pmatrix}\sqrt{\frac{\|\tilde{\mbf{R}}_f(\mbf{s}_f-\mbf{s}_f^{'})\|^2}{2\sigma^2}}\end{pmatrix}\end{bmatrix}\end{equation}
where $D_f$ is the $n_f$ dimensional subset of $D$.\end{prop}
\begin{IEEEproof}
$P_{err}\leq \sum_{f=1}^{k+2}P_{err,f}$ where $P_{err,f}$ is the error probability in searching $\mbf{s}_f$. And, $P_{err,f}\leq E_{\{\mbf{s}_f\in D_f\}}\begin{bmatrix}\sum_{\{\mbf{s}_{f}^{'}\in D_{f},\mbf{s}_{f}^{'}\neq\mbf{s}_{f}\}}\mathfrak{Q}\begin{pmatrix}\sqrt{\frac{\|\tilde{\mbf{R}}_{f}(\mbf{s}_{f}-\mbf{s}_{f}^{'})\|^2}{2\sigma^2}}\end{pmatrix}\end{bmatrix}$ by the above argument.
\end{IEEEproof}
To find the average error probability or its bound, when the channel matrix is not fixed but has some specific properties, we need the following lemma.
\begin{lemma}
\label{lem1}
Let $\mbf{H}$ be an $m\times n$ ($m\geq n$) random matrix with independently distributed columns such that each column has a distribution that is rotationally invariant from the left i.e. for any $m\times m$ unitary matrix $\mbf{\Theta}$ the distribution of $i$th column, $\mbf{H}[1:m][i:i]$, is equal to the distribution of $\mbf{\Theta}\mbf{H}[1:m][i:i]$. Then $\mbf{Q}$ and $\mbf{R}$, which constitute a thin QR decomposition $\mbf{H}=\mbf{QR}$ with the diagonal entries of $\mbf{R}$ non-negative, satisfy the following:
\begin{enumerate}
\item $\mbf{Q}$ and $\mbf{R}$ are independent random matrices.
\item The distribution of $\mbf{Q}$ is invariant under left-multiplication by any $m\times m$ unitary matrix, i.e., $\mbf{Q}$ has an isotropic distribution.
\item Considering the split form (\ref{e5}) and the notation of $\tilde{\mbf{R}}_f=\mbf{R}[i_{f-2}+1:i_{f-1}][i_{f-2}+1:i_{f-1}]$, for each $1\leq f \leq k+2$, $\tilde{\mbf{R}}_f$ has the same distribution as the upper triangular matrix obtained from the QRD of $\mbf{H}_f$ and $\tilde{\mbf{R}}_f^{*}\tilde{\mbf{R}}_f$ has the same distribution as $\mbf{H}_f^{*}\mbf{H}_f$ where $\mbf{H}_f=\mbf{H}[i_{f-2}+1:m][i_{f-2}+1:i_{f-1}]$: i.e. \begin{equation}\label{e16}\begin{split}&\tilde{\mbf{R}}_1^{*}\tilde{\mbf{R}}_1\stackrel{d}{=}\mbf{H}[1:m][1:i_0]^{*}\mbf{H}[1:m][1:i_0]\\
    &\tilde{\mbf{R}}_2^{*}\tilde{\mbf{R}}_2\stackrel{d}{=}\mbf{H}[i_0+1:m][i_0+1:i_1]^{*}\mbf{H}[i_0+1:m][i_0+1:i_1]\\
    &\mbox{\ \ \ \ }\vdots\\
    &\tilde{\mbf{R}}_f^{*}\tilde{\mbf{R}}_f\stackrel{d}{=}\mbf{H}[i_{f-2}+1:m][i_{f-2}+1:i_{f-1}]^{*}\mbf{H}[i_{f-2}+1:m][i_{f-2}+1:i_{f-1}]\\
    &\mbox{\ \ \ \ }\vdots\\
    &\tilde{\mbf{R}}_{k+2}^{*}\tilde{\mbf{R}}_{k+2}\stackrel{d}{=}\mbf{H}[i_{k}+1:m][i_{k}+1:n]^{*}\mbf{H}[i_{k}+1:m][i_{k}+1:n]\end{split}\end{equation} where $\mbf{A}\stackrel{d}{=}\mbf{B}$ denotes that $\mbf{A}$ has the same distribution as $\mbf{B}$.
\end{enumerate}
\end{lemma}
\begin{IEEEproof}
The proof of this lemma stems from the proof of Lemma 1 of \cite{Ha1} and the results of \cite{Edd}, and to prove item 3) we add some process and statements.

$\mbf{Q}$ is the partial matrix composed of the first $n$ columns of an $m\times m$ unitary matrix $\mbf{Q}_0$ where $\mbf{H}=\mbf{Q}_0\mbf{R}$ is a full version of QRD of $\mbf{H}$. $\mbf{Q}=\mbf{Q}_0[1:m][1:n]$. $\mbf{Q}_0$ and $\mbf{R}$ are independent and $\mbf{Q}_0$ is isotropically distributed, by Lemma 1 of \cite{Ha1}. Thus 1) and 2) are immediately followed.

Since the columns of $\mbf{H}$ are independent, the probability that $\mbf{H}$ has full column rank is 1. The columns of any sub-matrix $\mbf{G}$ of $\mbf{H}$ are independent and $\mbf{G}$ has full column rank with probability 1. Therefore, the upper triangular matrix with nonnegative diagonal entries which constitutes QRD of $\mbf{G}$ is unique and the thin QRD of $\mbf{G}$ with the diagonal entries of the upper triangular matrix nonnegative is unique, where $\{\mbf{G}\}\ni\mbf{H}$. From now on the diagonal entries of the triangular matrix of a QRD are non-negative.
Let $\mbf{H}_1=\mbf{H}[1:m][1:i_0]$ be QR decomposed as \[\mbf{H}_1=\mbf{Q}_1\begin{bmatrix}\mbf{T}_1\\0\end{bmatrix}\] where $\mbf{Q}_1$ is $m\times m$ unitary and $\mbf{T}_1$ $i_0\times i_0$ upper triangular. Applying $\mbf{Q}_1^{*}$ to the full $\mbf{H}$ we have \[\mbf{Q}_1^{*}\mbf{H}=\begin{bmatrix} \mbf{T}_1 && \mbf{A}_1\\0 && \tilde{\mbf{H}}_1\end{bmatrix}\] where $\begin{bmatrix}\mbf{A}_1\\\tilde{\mbf{H}}_1\end{bmatrix}=\mbf{Q}_1^{*}\mbf{H}[1:m][i_0+1:n]$. $\begin{bmatrix}\mbf{A}_1\\\tilde{\mbf{H}}_1\end{bmatrix}$ is independent of $\mbf{Q}_1$ and \[\begin{bmatrix}\mbf{A}_1\\\tilde{\mbf{H}}_1\end{bmatrix}\stackrel{d}{=}\mbf{H}[1:m][i_0+1:n]\] by the rotational invariance of the columns of $\mbf{H}$. Thus $\tilde{\mbf{H}}_1\stackrel{d}{=}\mbf{H}[i_0+1:m][i_0+1:n]$ and $\tilde{\mbf{H}}_1[1:m-i_0][1:n_2]\stackrel{d}{=}\mbf{H}_2$, recalling $n_f=i_{f-1}-i_{f-2}$. Let $\tilde{\mbf{H}}_1[1:m-i_0][1:n_2]$ be QR decomposed as \[\tilde{\mbf{H}}_1[1:m-i_0][1:n_2]=\mbf{Q}_2\begin{bmatrix}\mbf{T}_2\\0\end{bmatrix}\] where $\mbf{Q}_2$ is $(m-i_0)\times(m-i_0)$ unitary and $\mbf{T}_2$ $n_2\times n_2$ upper triangular. Then we have \[\mbf{Q}_2^{*}\tilde{\mbf{H}}_1=\mbf{Q}_2\begin{bmatrix}\mbf{T}_2 && \mbf{A}_2\\ 0 && \tilde{\mbf{H}}_2\end{bmatrix}\] where \begin{equation*}\begin{split}\begin{bmatrix}\mbf{A}_2\\ \tilde{\mbf{H}}_2\end{bmatrix}&=\mbf{Q}_2^{*}\tilde{\mbf{H}}_1[1:m-i_0][n_2+1:n-i_0]\\&\stackrel{d}{=}\tilde{\mbf{H}}_1[1:m-i_0][n_2+1:n-i_0].
\end{split}\end{equation*} Hence $\tilde{\mbf{H}}_2\stackrel{d}{=}\mbf{H}[i_1+1:m][i_1+1:n]$ and  $\tilde{\mbf{H}}_2[1:m-i_1][1:n_3]\stackrel{d}{=}\mbf{H}_3$. For $3\leq f \leq k+2$, $\tilde{\mbf{H}}_{f-1}\stackrel{d}{=}\mbf{H}[i_{f-2}+1:m][i_{f-2}+1:n]$ and $\tilde{\mbf{H}}_{f-1}[1:m-i_{f-2}][1:n_f]\stackrel{d}{=}\mbf{H}_f$. $\tilde{\mbf{H}}_{f-1}[1:m-i_{f-2}][1:n_f]$ is QR decomposed as \[\tilde{\mbf{H}}_{f-1}[1:m-i_{f-2}][1:n_f]=\mbf{Q}_f\begin{bmatrix}\mbf{T}_f\\0\end{bmatrix}\] where $\mbf{Q}_f$ is $(m-i_{f-2})\times(m-i_{f-2})$ unitary and $\mbf{T}_f$ is $n_f\times n_f$ upper triangular.
Now, we have \begin{equation*}\begin{split}\mbf{H}&=\mbf{Q}_1\begin{bmatrix}\mbf{T}_1 && \mbf{A}_1\\ 0 && \tilde{\mbf{H}}_1\end{bmatrix}\\&=\mbf{Q}_1\begin{bmatrix}\mbf{I}_{n_1} && 0 \\ 0 && \mbf{Q}_2\end{bmatrix}\begin{bmatrix}\mbf{T}_1 && \mbf{A}_1\\ 0 && \begin{bmatrix}\mbf{T}_2 && \mbf{A}_2\\ 0 && \tilde{\mbf{H}}_2\end{bmatrix}\end{bmatrix}\\
&=\mbf{Q}_1\begin{bmatrix}\mbf{I}_{n_1} && 0 \\ 0 && \mbf{Q}_2\end{bmatrix}\cdots\begin{bmatrix}\mbf{I}_{n_1} && 0 && 0 && 0\\ 0 && \mbf{I}_{n_2} && 0 && 0\\ \vdots && \ddots && \vdots && \vdots\\0 && 0 && \mbf{I}_{n_{k+1}} && 0 \\0 && \cdots && 0 && \mbf{Q}_{k+2}\end{bmatrix}\begin{bmatrix}\mbf{T}_1 && \mbf{A}_1 \\ 0 && \begin{bmatrix}\mbf{T}_2 && \mbf{A}_2 \end{bmatrix}\\ \vdots && \ddots \\ 0 && \begin{bmatrix} 0 &&\mbf{T}_{k+2} \\ 0 && 0 \end{bmatrix}\end{bmatrix}.\end{split}\end{equation*}
We have, with probability 1, \begin{equation}\label{e17}\mbf{R}=\begin{bmatrix}\mbf{T}_1 && \mbf{A}_1 \\ 0 && \begin{bmatrix}\mbf{T}_2 && \mbf{A}_2 \end{bmatrix}\\ \vdots && \ddots \\ 0 && \begin{bmatrix} 0 &&\mbf{T}_{k+2}\end{bmatrix}\end{bmatrix}\end{equation} and $\tilde{\mbf{R}}_f=\mbf{T}_f$ for all $1\leq f \leq k+2$. By the rotational invariance, this concludes the third statement.
\end{IEEEproof}
Even when sorting columns intervenes during QR-decomposition, Lemma \ref{lem1} is verified. Now, if $\mbf{H}$ is a random matrix satisfying the condition of Lemma \ref{lem1}, we have $E_{\mbf{H}}\begin{bmatrix}\mathfrak{Q}\begin{pmatrix}\sqrt{\frac{\|\tilde{\mbf{R}}_f(\mbf{s}_f-\mbf{s}_f^{'})\|^2}{2\sigma^2}}\end{pmatrix}\end{bmatrix}=
E_{\mbf{H}}\begin{bmatrix}\mathfrak{Q}\begin{pmatrix}\sqrt{\frac{\|\mbf{H}_f(\mbf{s}_f-\mbf{s}_f^{'})\|^2}{2\sigma^2}}\end{pmatrix}\end{bmatrix}$ and the following result.
\begin{theorem}\label{thm1}If random matrix $\mbf{H}$ is under the condition of Lemma \ref{lem1} then the average error probability $P_{err}$ for the divided decoding (\ref{e5}) satisfies
\begin{equation}\label{e18} \begin{split}P_{err}&\leq E_{\mbf{H}}\begin{bmatrix}\sum_{f=1}^{k+2}E_{\{\mbf{s}_f\in D_f\}}\begin{bmatrix}\sum_{\{\mbf{s}_f^{'}\in D_f,\mbf{s}_f^{'}\neq\mbf{s}_f\}} \mathfrak{Q}\begin{pmatrix}\sqrt{\frac{\|\mbf{H}_f(\mbf{s}_f-\mbf{s}^{'}_f)\|^2}{2\sigma^2}}\end{pmatrix}\end{bmatrix}\end{bmatrix}\\
&=\sum_{f=1}^{k+2}E_{\{\mbf{s}_f\in D_f\}}\begin{bmatrix}\sum_{\{\mbf{s}_f^{'}\in D_f,\mbf{s}_f^{'}\neq\mbf{s}_f\}}E_{\mbf{H}_f}\begin{bmatrix} \mathfrak{Q}\begin{pmatrix}\sqrt{\frac{\|\mbf{H}_f(\mbf{s}_f-\mbf{s}^{'}_f)\|^2}{2\sigma^2}}\end{pmatrix}\end{bmatrix}\end{bmatrix},
\end{split}\end{equation}
and if $\mbf{h}_f:=vec(\mbf{H}_f)$ \footnote{$vec(\mbf{H})$ of $m\times n$ matrix $\mbf{H}$ is defined as $\begin{bmatrix}\mbf{H}[:][1]\\\vdots\\\mbf{H}[:][n]\end{bmatrix}$ where $\mbf{H}[:][i]$ is the i-th column of $\mbf{H}$. } has a multi-dimensional complex normal distribution with mean $\mbf{0}$ and covariance matrix $\mbf{\Upsilon}_f$, i.e. $\mbf{h}_f\sim N_C(\mbf{0}, \mbf{\Upsilon}_f)$, then \begin{equation}\label{e19}\begin{split}E_{\mbf{H}_f}\begin{bmatrix}\mathfrak{Q}\begin{pmatrix}\sqrt{\frac{\|\mbf{H}_f(\mbf{s}_f-\mbf{s}^{'}_f)\|^2}{2\sigma^2}}\end{pmatrix}\end{bmatrix}&\leq
\big{|}\mbf{I}_{(m-i_{(f-2)})}+\frac1{4\sigma^2}(\mbf{S}_f-\mbf{S}_f^{'})\mbf{\Upsilon}_f(\mbf{S}_f-\mbf{S}_f^{'})^{*}\big{|}^{-1}\\ &\leq \big{(}\frac1{4\sigma^2}\big{)}^{-k_f}\prod\limits_{i=1}^{k_f}\epsilon_{(f,i)}^{-1}\end{split}\end{equation} where $\mbf{S}_f:=\mbf{s}_f^T\otimes\mbf{I}_{(m-i_{(f-2)})}$, $\mbf{S}_f^{'}:=(\mbf{s}_f^{'})^T\otimes\mbf{I}_{(m-i_{(f-2)})}$, $k_f:=\text{rank}\{(\mbf{S}_f-\mbf{S}_f^{'})\mbf{\Upsilon}_f(\mbf{S}_f-\mbf{S}_f^{'})^{*}\}$, $\{\epsilon_{(f,1)}, \dots, \epsilon_{(f,k_f)}\}$ are the nonzero eigenvalues of $(\mbf{S}_f-\mbf{S}_f^{'})\mbf{\Upsilon}_f(\mbf{S}_f-\mbf{S}_f^{'})^{*}$, $(\cdot)^T$ denotes the transpose, $\otimes$ means the Kronecker product, and $|\cdot|$ the determinant of a matrix.
\end{theorem}
\begin{IEEEproof} First, (\ref{e18}) is proved as follows:
\begin{equation*}\begin{split}P_{err}&\leq E_{\mbf{H}}\begin{bmatrix}\sum_{f=1}^{k+2}E_{\{\mbf{s}_f\in D_f\}}\begin{bmatrix}\sum_{\{\mbf{s}_f^{'}\in D_f,\mbf{s}_f^{'}\neq\mbf{s}_f\}}\mathfrak{Q}\begin{pmatrix}\sqrt{\frac{\|\tilde{\mbf{R}}_f(\mbf{s}_f-\mbf{s}_f^{'})\|^2}{2\sigma^2}}\end{pmatrix}\end{bmatrix}\end{bmatrix} \text{\ \ (by Proposition \ref{prop1})}\\&=E_{\mbf{H}}\begin{bmatrix}\sum_{f=1}^{k+2}E_{\{\mbf{s}_f\in D_f\}}\begin{bmatrix}\sum_{\{\mbf{s}_f^{'}\in D_f,\mbf{s}_f^{'}\neq\mbf{s}_f\}}\mathfrak{Q}\begin{pmatrix}\sqrt{\frac{\|\mbf{H}_f(\mbf{s}_f-\mbf{s}_f^{'})\|^2}{2\sigma^2}}\end{pmatrix}\end{bmatrix}\end{bmatrix} \text{\ \  (by Lemma \ref{lem1})}\\&=\sum_{f=1}^{k+2}E_{\{\mbf{s}_f\in D_f\}}\begin{bmatrix}\sum_{\{\mbf{s}_f^{'}\in D_f,\mbf{s}_f^{'}\neq\mbf{s}_f\}}E_{\mbf{H}_f}\begin{bmatrix} \mathfrak{Q}\begin{pmatrix}\sqrt{\frac{\|\mbf{H}_f(\mbf{s}_f-\mbf{s}^{'}_f)\|^2}{2\sigma^2}}\end{pmatrix}\end{bmatrix}\end{bmatrix}.\end{split}\end{equation*}
Secondly, adopting the approach of \cite{Lar}, by the Chernoff bound we have $\mathfrak{Q}\begin{pmatrix}\sqrt{\frac{\|\mbf{H}_f(\mbf{s}_f-\mbf{s}^{'}_f)\|^2}{2\sigma^2}}\end{pmatrix}\leq \exp{\Big{(}-\frac{\|\mbf{H}_f(\mbf{s}_f-\mbf{s}^{'}_f)\|^2}{4\sigma^2}\Big{)}}=\exp{\Big{(}-\frac{\|(\mbf{S}_f-\mbf{S}^{'}_f)\mbf{h}_f\|^2}{4\sigma^2}\Big{)}}$. The covariance matrix of $(\mbf{S}_f-\mbf{S}^{'}_f)\mbf{h}_f$ is $(\mbf{S}_f-\mbf{S}_f^{'})\mbf{\Upsilon}_f(\mbf{S}_f-\mbf{S}_f^{'})^{*}$ and $\|(\mbf{S}_f-\mbf{S}^{'}_f)\mbf{h}_f\|^2=\sum_{i=1}^{k_f}\chi_i^2$, where $\{\chi_i^2\}_{i=1}^{k_f}$ are independent and the density function $p_{\chi_i^2}(x)$ of $\chi_i^2$ is $\frac1{\epsilon_{(f,i)}}\exp{\big{(}-\frac{x}{\epsilon_{(f,i)}}\big{)}}$. Hence we get \begin{equation}\begin{split}E_{\mbf{H}_f}\begin{bmatrix}\mathfrak{Q}\begin{pmatrix}\sqrt{\frac{\|\mbf{H}_f(\mbf{s}_f-\mbf{s}^{'}_f)\|^2}{2\sigma^2}}\end{pmatrix}\end{bmatrix}&\leq
E_{\mbf{h}_f}\begin{bmatrix}\exp{\Big{(}-\frac{\|(\mbf{S}_f-\mbf{S}^{'}_f)\mbf{h}_f\|^2}{4\sigma^2}\Big{)}}\end{bmatrix}\\
&=\int_0^{\infty}\cdots\int_0^{\infty}\prod_{i=1}^{k_f}\begin{pmatrix}\frac1{\epsilon_{(f,i)}}\exp{\Big{(}-\frac{t_i^2}{4\sigma^2}-\frac{t_i^2}{\epsilon_{(f,i)}}\Big{)}}\end{pmatrix}dt_1^2\cdots dt^2_{k_f}\\&=\prod_{i=1}^{k_f}\begin{pmatrix}\frac1{\epsilon_{(f,i)}}\int_0^{\infty} dt^2\exp{\Big{(}-\frac{t^2}{4\sigma^2}-\frac{t^2}{\epsilon_{(f,i)}}\Big{)}}\end{pmatrix}\\&=\prod_{i=1}^{k_f}\Big{(}\frac{\epsilon_{(f,i)}}{4\sigma^2}+1\Big{)}^{-1}\\
&=\Big{|}\mbf{I}_{k_f}+diag(\epsilon_{(f,1)}, \dots, \epsilon_{(f,k_f)})\Big{|}^{-1}\\
&=\Big{|}\mbf{I}_{(m-i_{(f-2)})}+\frac1{4\sigma^2}(\mbf{S}_f-\mbf{S}_f^{'})\mbf{\Upsilon}_f(\mbf{S}_f-\mbf{S}_f^{'})^{*}\Big{|}^{-1}.\end{split}\end{equation}
Since $\Big{(}\frac{\epsilon_{(f,i)}}{4\sigma^2}+1\Big{)}^{-1}\leq\Big{(}\frac{\epsilon_{(f,i)}}{4\sigma^2}\Big{)}^{-1}$, the second inequality of (\ref{e19}) is obviously true.
\end{IEEEproof}

Proposition \ref{prop1} and Theorem \ref{thm1} can be generalized when we use a divided decoding to find $n\times r$ matrix $\mbf{X}$ from $m\times r$ matrix $\mbf{Y}$ such that \begin{equation}\label{e20}\mbf{Y}=\mbf{H}\mbf{X}+\mbf{E},\end{equation} where the entries of $\mbf{E}$ are independent complex Gaussian random variables with mean zero and variance $\sigma^2$. Let $\mbf{X}_f:=\mbf{X}[i_{f-2}+1:i_{f-1}][1:r]$, $\mathbb{D}$ be the domain that $\mbf{X}$ belongs to, $\mathbb{D}_f$ the domain that $\mbf{X}_f$ belongs to.
\begin{prop}
\label{prop2}
For fixed $\mbf{H}$, the error probability $P_{err}$ in detecting $\mbf{X}$ by using divided decoding with each sub-matrix $\mbf{X}_f$ found by a detector searching ML point satisfies that \begin{equation}\label{e21} P_{err}\leq\sum_{f=1}^{k+2}E_{\{\mbf{X}_f\in\mathbb{D}_f\}}\begin{bmatrix}\sum_{\{\mbf{X}_f^{'}\in\mathbb{D}_f,\mbf{X}_f^{'}\neq\mbf{X}_f\}}
\mathfrak{Q}\begin{pmatrix}\sqrt{\frac{\|\tilde{\mbf{R}}_f(\mbf{X}_f-\mbf{X}_f^{'}\|^2}{2\sigma^2}}\end{pmatrix}\end{bmatrix}\end{equation}
\end{prop}
\begin{IEEEproof}
$P_{err}\leq \sum_{f=1}^{k+2}P_{err,f}$ where $P_{err,f}$ is the error probability in searching $\mbf{X}_f$. And the remainder is similar to that of Proposition \ref{prop1}.
\end{IEEEproof}
\begin{theorem}
\label{thm2}
If random matrix $\mbf{H}$ is under the condition of Lemma \ref{lem1} then the average error probability $P_{err}$ for the divided decoding (\ref{e5}) satisfies \begin{equation}\label{e22}P_{err}\leq \sum_{f=1}^{k+2}E_{\{\mbf{X}_f\in\mathbb{D}_f\}}\begin{bmatrix}\sum_{\{\mbf{X}^{'}_f\in\mathbb{D}_f, \mbf{X}^{'}_f\neq\mbf{X}_f\}}E_{\mbf{H}_f}\begin{bmatrix}\mathfrak{Q}\begin{pmatrix}\sqrt{\frac{\|\mbf{H}_f(\mbf{X}_f-\mbf{X}_f^{'})\|^2}{2\sigma^2}}\end{pmatrix}\end{bmatrix}\end{bmatrix},
\end{equation}
and if $\mbf{h}_f=vec(\mbf{H}_f)\sim N_C(\mbf{0},\mbf{\Upsilon}_f)$ then \begin{equation}\label{e23}\begin{split}&E_{\mbf{H}_f}\begin{bmatrix}\mathfrak{Q}\begin{pmatrix}\sqrt{\frac{\|\mbf{H}_f(\mbf{X}_f-\mbf{X}_f^{'})\|^2}{2\sigma^2}}\end{pmatrix}\end{bmatrix}\leq\\ &\big{|}\mbf{I}_{r(m-i_{(f-2)})}+\frac1{4\sigma^2}[(\mbf{X}_f-\mbf{X}_f^{'})^T\otimes\mbf{I}_{(m-i_{(f-2)})}]\mbf{\Upsilon}_f[(\mbf{X}_f-\mbf{X}_f^{'})^T\otimes\mbf{I}_{(m-i_{(f-2)})}]^{*}\big{|}^{-1}\\
&\leq\Big{(}\frac1{4\sigma^2}\Big{)}^{-k_f}\prod_{i=1}^{k_f}\epsilon_{(f,i)}^{-1}\end{split}\end{equation} where $k_f:=\text{rank}\{[(\mbf{X}_f-\mbf{X}_f^{'})^T\otimes\mbf{I}_{(m-i_{(f-2)})}]\mbf{\Upsilon}_f[(\mbf{X}_f-\mbf{X}_f^{'})^T\otimes\mbf{I}_{(m-i_{(f-2)})}]^{*}\}$, and $\{\epsilon_{(f,1)}, \dots, \epsilon_{(f,k_f)}\}$ are the nonzero eigenvalues of $[(\mbf{X}_f-\mbf{X}_f^{'})^T\otimes\mbf{I}_{(m-i_{(f-2)})}]\mbf{\Upsilon}_f[(\mbf{X}_f-\mbf{X}_f^{'})^T\otimes\mbf{I}_{(m-i_{(f-2)})}]^{*}$. \end{theorem}
\begin{IEEEproof} The proof is a simple extension of the proof of Theorem \ref{thm1}.
\end{IEEEproof}
If we assume $vec(\mbf{H})\sim N_C(0,\rho^2\mbf{I}_{mn})$, then $\mbf{\Upsilon}_f=\rho^2\mbf{I}_{d_f}$ where $d_f=(m-i_{(f-2)})(i_{(f-1)}-i_{(f-2)})$ and we have  \begin{equation}\label{e24}\begin{split}&E_{\mbf{H}_f}\begin{bmatrix}\mathfrak{Q}\begin{pmatrix}\sqrt{\frac{\|\mbf{H}_f(\mbf{X}_f-\mbf{X}_f^{'})\|^2}{2\sigma^2}}\end{pmatrix}\end{bmatrix}\\
&\leq\big{|}\mbf{I}_{r(m-i_{(f-2)})}+\frac{\rho^2}{4\sigma^2}[(\mbf{X}_f-\mbf{X}_f^{'})^T\otimes\mbf{I}_{m-i_{(f-2)}}]\mbf{I}_{d_f}[(\mbf{X}_f-\mbf{X}_f^{'})^T\otimes\mbf{I}_{m-i_{(f-2)}}]^{*}\big{|}^{-1}\\&=
\big{|}\mbf{I}_r+\frac{\rho^2}{4\sigma^2}(\mbf{X}_f-\mbf{X}_f^{'})^{*}(\mbf{X}_f-\mbf{X}_f^{'})\big{|}^{(-m+i_{(f-2)})}\\
&=\big{|}\mbf{I}_r+\frac{\rho^2}{4\sigma^2}(\mbf{X}_f-\mbf{X}_f^{'})(\mbf{X}_f-\mbf{X}_f^{'})^{*}\big{|}^{(-m+i_{(f-2)})}\\
&\leq\big{|}(\mbf{X}_f-\mbf{X}_f^{'})(\mbf{X}_f-\mbf{X}_f^{'})^{*}\big{|}^{(-m+i_{(f-2)})}\cdot \Big{(}\frac{\rho^2}{4\sigma^2}\Big{)}^{-d_f}.\end{split}\end{equation}
Hence, we get $P_{err,f}\leq\Big{(}\frac{\rho^2}{4\sigma^2}\Big{)}^{-d_f}\cdot G_f$ where \[G_f=E_{\{\mbf{X}_f\in\mathbb{D}_f\}}\begin{bmatrix}\sum_{\{\mbf{X}^{'}_f\in\mathbb{D}_f, \mbf{X}^{'}_f\neq\mbf{X}_f\}}\big{|}(\mbf{X}_f-\mbf{X}_f^{'})(\mbf{X}_f-\mbf{X}_f^{'})^{*}\big{|}^{(-m+i_{(f-2)})}\end{bmatrix}\] and the diversity order of $P_{err,f}$ is $d_f$. The diversity order of $P_{err}=\sum_{f=1}^{k+2}P_{err,f}$ is a combination of $\{d_f\}_{f=1}^{k+2}$.

When (\ref{e3}) (or (\ref{e11}), more generally (\ref{e20})) is split according to both $\{i_0, \dots, i_k\}$ and $\{j_0, \dots, j_k\}$ and all sub-vectors detected by a ML decoder, for example SD; even if the set of the sub-vector sizes are equal i.e. $\{(i_{f-1}-i_{f-2})\}_{f=1}^{k+2}=\{(j_{f-1}-j_{f-2})\}_{f=1}^{k+2}$, the diversity-orders and error-rates of the two are different and significantly different for many cases. On the other hand, the complexities of the two are not so different, which will be explained with simulation results in the next section.
\begin{ex}\label{ex1}Consider the example of $\{i_0=1\}$ and $\{j_0=n-1\}$, where $vec(\mbf{H})\sim N_C(0,\rho^2\mbf{I}_{mn})$ and the sets of sub-vector sizes of these two are equal to $\{1, n-1\}$. But then we have \[P_{err}(\{i_0=1\})\leq G_1(i_0)\cdot\Big{(}\frac{\rho^2}{4\sigma^2}\Big{)}^{-m}+G_2(i_0)\cdot\Big{(}\frac{\rho^2}{4\sigma^2}\Big{)}^{-(m-1)(n-1)}\] and \[P_{err}(\{j_0=n-1\}\leq G^{'}_1(j_0)\cdot\Big{(}\frac{\rho^2}{4\sigma^2}\Big{)}^{-m(n-1)}+G^{'}_2(j_0)\cdot\Big{(}\frac{\rho^2}{4\sigma^2}\Big{)}^{-m+n-1}\] where \begin{equation*}\begin{split}G_1(i_0)&=E_{\{\mbf{s}_1\}}\begin{bmatrix}\sum_{\{\mbf{s}^{'}_1, \mbf{s}^{'}_1\neq\mbf{s}_1\}}\big{|}(\mbf{s}_1-\mbf{s}_1^{'})(\mbf{s}_1-\mbf{s}_1^{'})^{*}\big{|}^{-m}\end{bmatrix},\\ G_2(i_0)&=E_{\{\mbf{s}_2\}}\begin{bmatrix}\sum_{\{\mbf{s}^{'}_2, \mbf{s}^{'}_2\neq\mbf{s}_2\}}\big{|}(\mbf{s}_2-\mbf{s}_2^{'})(\mbf{s}_2-\mbf{s}_2^{'})^{*}\big{|}^{-m+1}\end{bmatrix},\\ G_1^{'}(j_0)&=E_{\{\mbf{s}_2\}}\begin{bmatrix}\sum_{\{\mbf{s}^{'}_2, \mbf{s}^{'}_2\neq\mbf{s}_2\}}\big{|}(\mbf{s}_2-\mbf{s}_2^{'})(\mbf{s}_2-\mbf{s}_2^{'})^{*}\big{|}^{-m}\end{bmatrix},\\ G_2^{'}(j_0)&=E_{\{\mbf{s}_1\}}\begin{bmatrix}\sum_{\{\mbf{s}^{'}_1, \mbf{s}^{'}_1\neq\mbf{s}_1\}}\big{|}(\mbf{s}_1-\mbf{s}_1^{'})(\mbf{s}_1-\mbf{s}_1^{'})^{*}\big{|}^{-m+n-1}\end{bmatrix},\\  &\mbf{s}_1=\mbf{s}[1:1], \mbf{s}_2=\mbf{s}[2:n].\end{split}\end{equation*} This example shows that $P_{err}(\{i_0\})$ has larger diversity order and is at the same time much lower than $P_{err}(\{j_0\})$ if $m, n > 2$.
\end{ex}
Example \ref{ex1} is a simplest comparison, whose generalized version can be obtained for the pair of $\{i_0=l\}$ and $\{j_0=n-l\}$ and more expansively for a class of sets of the form $\{i_0, \dots, i_k\}$ whose resultant sets of sub-vector sizes are identical. From this reasoning we have the following conjecture.
\begin{con}\label{con1}
If $\hat{\mbf{s}}$, $\hat{\mbf{s}}_M$, or $\hat{\mbf{s}}_{\epsilon}$ is approximated by divided decoding with SD according to splitting index set $\{i_0, i_1, \dots, i_k\}$ ($1\leq i_0<i_1<\cdots<i_k<n$) whose sub-vector size set is fixed as $\{n_f\}_{f=1}^{k+2}$, $n_f=i_{f-1}-i_{f-2}$, then the index set $\{i_0, i_1, \dots, i_k\}$ letting $\{n_f\}$ be $n_1\leq n_2\leq \cdots\leq n_{k+2}$ is the best choice, i.e. it makes the error rate and the complexity least at the same time.
\end{con} The reasoning of this choice letting the complexity least under fixed $\{n_f\}_{f=1}^{k+2}$ is that the error propagation from the sub-vectors previously found is least at each step of searching a present sub-vector solution by SD and the complexity of SD depends on SNR and the sub-vector size.

\subsection{complexity}
To see roughly the gain in complexity; if we use the full search algorithm then the number of multiplications required for the computation except QRD is $2n(n+3)q^n$ for $q-$QAM constellation, but if we apply a divided decoding which splits a signal vector into $k$ ones of equal size and detects each sub-vector by full search then the number of multiplications required is $2n\big{(}(\frac{n}k+3)q^{n/k}+\frac{n}k(k-1)\big{)}$. If we apply (\ref{e4}) with full search then the number of multiplications required is $2(n-i_0)(n-i_0+3)q^{n-i_0}+4i_0(n-i_0)+2i_0(i_0+3)q^{i_0}$. The exponent of $q$ depends on the sub-vector sizes. After QRD, if the mother search algorithm's complexity is $f(n)$ and depends only on $n$ then the complexity of the divided decoding with $k$ splits of equal size is $kf(n/k)+2n^2(k-1)/k$ ($kf(n/k)+n^2(k-1)/(2k)$ for real systems) and the complexity of applying (\ref{e4}) is $f(i_0)+f(n-i_0)+4i_0(n-i_0)$ ($f(i_0)+f(n-i_0)+i_0(n-i_0)$ for real systems). $kf(n/k)$ and $f(i_0)+f(n-i_0)$ multiplications are required for search, and $2n^2(k-1)/k$ and $4i_0(n-i_0)$ multiplications are for cancelling.

If a given search algorithm after QRD has its complexity $f(n)$ only dependent on the size $n$ of the vector searched then the complexity of divided decoding based on the search algorithm is obtained by simple calculation as follows:
\begin{prop}
The complexity of divided decoding according to splitting index set $\{i_0, i_1, \dots, i_k\}$ with sub-vector size set $\{n_j\}_{j=1}^{k+2}$, $n_j=i_{j-1}-i_{j-2}$, is $\sum_{j=1}^{k+2}f(n_j)+A\begin{pmatrix}(n_j)_{j=2}^{k+2}\end{pmatrix}$ where $A\begin{pmatrix}(n_j)_{j=2}^{k+2}\end{pmatrix}=4\sum_{j=2}^{k+2}n_ji_{j-2}$ for complex systems and $A\begin{pmatrix}(n_j)_{j=2}^{k+2}\end{pmatrix}=\sum_{j=2}^{k+2}n_ji_{j-2}$ for real systems, and in most cases $A\begin{pmatrix}(n_j)_{j=2}^{k+2}\end{pmatrix}\propto \sum_{j=2}^{k+2}n_ji_{j-2}$.
\end{prop}

If the complexity of a given search algorithm $\mathbb{A}$ after QRD depends on the statistical property of $\mbf{H}$, SNR, $m,n$ and particularly depends on $\{n, m, \sigma^2\}$ i.e. $f=f(n, m, \sigma^2)$ then we have the following formula:
\begin{theorem} \label{thmtttt}If random matrix $\mbf{H}$ is under the condition of Lemma \ref{lem1} then
the complexity of divided decoding according to splitting index set $\{i_0, i_1, \dots, i_k\}$, $f_d(n, m, \sigma^2)$, is $f_d(n, m, \sigma^2)=\sum_{j=1}^{k+2}f(n_j, m-i_{j-2}, \sigma^2)+A\begin{pmatrix}(n_j)_{j=2}^{k+2}\end{pmatrix}$.
\end{theorem}
\begin{IEEEproof}
For each $j$, $1\leq j\leq k+2$, divided decoding based on the search algorithm $\mathbb{A}$ finds $\hat{\mbf{s}}_{j}$ from the following equation \[\tilde{\mbf{y}}_j=\tilde{\mbf{R}}_j\mbf{s}_j+\mbf{z}_j.\] The elements of $\mbf{z}_j$ are i.i.d. with circularly symmetric complex normal variables with mean zero and variance $\sigma^2$. $\tilde{\mbf{R}}_j$ has the same distribution as the upper triangular matrix obtained from the QRD of $\mbf{H}_j=\mbf{H}[i_{j-2}+1:m][i_{j-2}+1:i_{j-1}]$ from Lemma \ref{lem1}. Therefore the complexity required for finding $\hat{\mbf{s}}_{j}$ is $f(n_j, m-i_{j-2}, \sigma^2)$. By summing up $f(n_j, m-i_{j-2}, \sigma^2)$ over $j$ and $ A\begin{pmatrix}(n_j)_{j=2}^{k+2}\end{pmatrix}$, $f_d(n, m, \sigma^2)=\sum_{j=1}^{k+2}f(n_j, m-i_{j-2}, \sigma^2)+A\begin{pmatrix}(n_j)_{j=2}^{k+2}\end{pmatrix}$.
\end{IEEEproof}

\begin{figure}[!t]
\centering
\includegraphics[width=2.8in, clip]{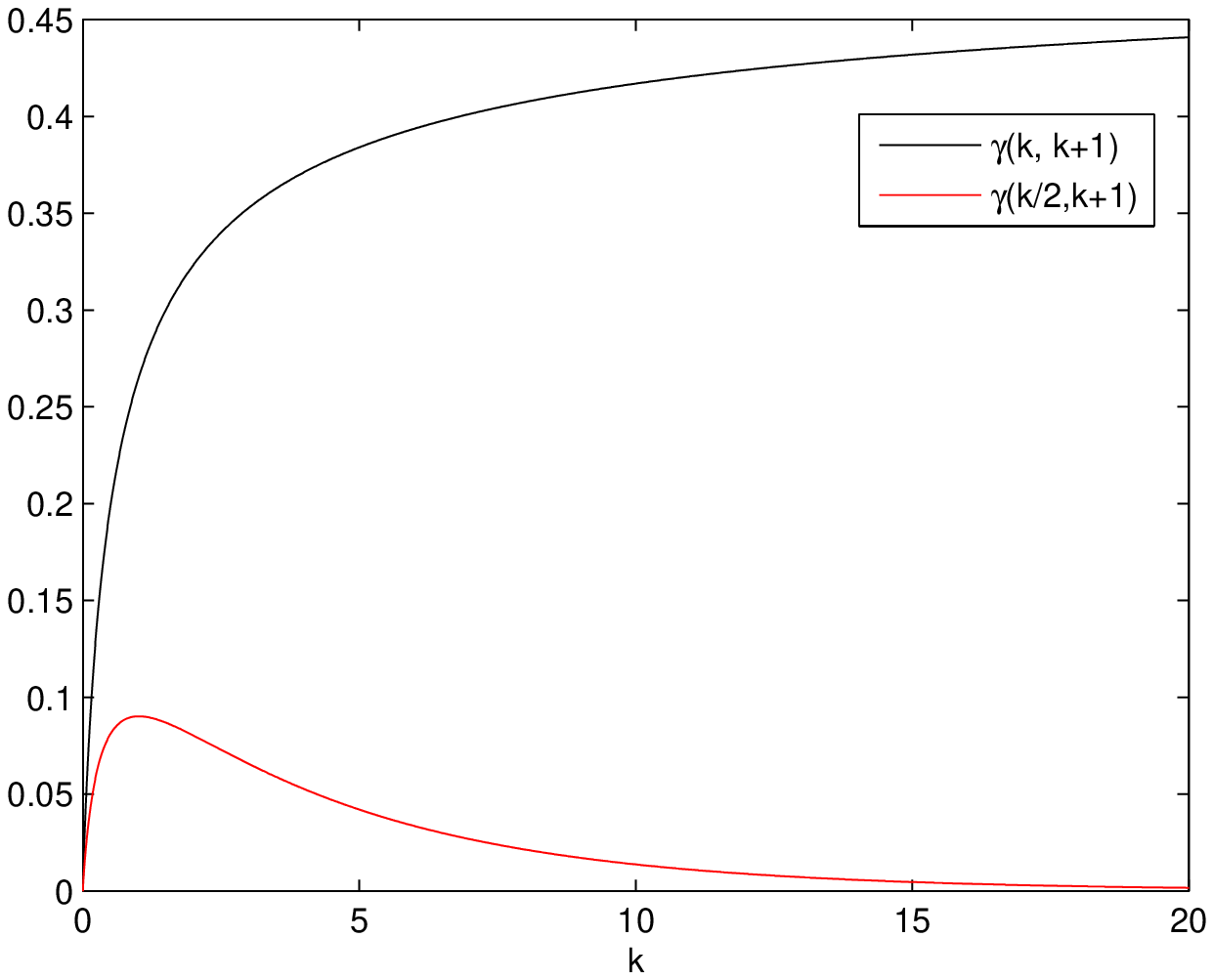}
\caption{The behavior of $\gamma(k,k+1)$ and $\gamma(k/2, k+1)$} \label{fig1}
\end{figure}

The expected complexity for SD of Finke and Pohst under Rayleigh channel estimated in \cite{Ha1} depends on $\{n, m, \sigma^2\}$. (We omit the constellation number which also have an effect on the expected complexity since we only focus on the alterations and effects via divided decoding.) From the estimated formula the expected complexity of SD is more dependent on $n$ than $m$. The expected complexity of SD with Schnorr-Euchner's strategy is known to be less than Finke and Pohst's in practical experiment because the Schnorr-Euchner's starts with closer point to the ML point. The sphere radius determined by the first point (which is the ZF-SIC solution) of Schnorr-Euchner's search is efficient because it does not need any extra calculation. The estimation in \cite{Ha1} is an upper bound of the expected complexities for SD of Schnorr-Euchner and other advanced SD's. The expected complexity calculated in \cite{Ha1} is a summation of terms taking the form of a combinatorial number multiplied by $\gamma(a, (m-n+k)/2)=\int_0^a t^{(m-n+k)/2-1}e^{-t}dt/\Gamma((m-n+k)/2)$ where $k$ varies from 1 to $n$ and $a$ depends on $m$ and SNR. The expected complexity of SD grows exponentially in $n$ but the formula proposed in \cite{Ha1} describes that the complexity is approximately cubic in $n$ for mid to high SNR and some range of $m$ and $n$. It is hard to find the form of the largest value of $a$ such that $\gamma(a, (m-n+k)/2)$ decreases as $(m-n+k)/2)$ increases and can be ignored for $(m-n+k)/2)> k_0$ for a proper value $k_0$. But, we can find out roughly the behavior of $\gamma(a, (m-n+k)/2)$ as follows: as shown in \figurename \ref{fig1}, when $a=(m-n+k)/2-1$ then the value of $\gamma(a, (m-n+k)/2)$ increases as $(m-n+k)/2)$ does, $\gamma(a, (m-n+k)/2)>0.1$ and can not be ignored. But when $a=\frac1{2}((m-n+k)/2-1)$ then $\gamma(a, (m-n+k)/2)$ decreases as $(m-n+k)/2)$ increases for $(m-n+k)/2)>2$ and $\gamma(a, (m-n+k)/2)<0.05$ for $(m-n+k)/2>4$.  $a$ is proportional to $m/(1+\beta\cdot SNR)$ for some constant $\beta$, and the number of constituent terms of the expected complexity strongly depends on $n$. The dependency of SD's complexity on $m$ is much less than the size $n$ of the vector searched. For mid to low SNR range, the slope of complexity versus SNR is very steep for $n \geq 4$ and increases as $n$ does. Divided decoding mitigates the slope increase since the combinatorial terms are summed up only within the sizes of sub-vectors to obtain the complexity.

\section{Simulation Results} \label{sec5}

\begin{figure}[!t]
\centering
\includegraphics[width=2.8in, clip]{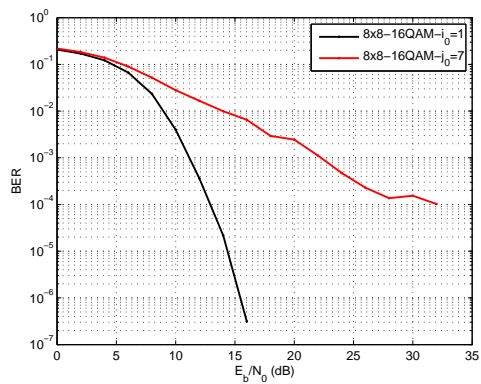}
\caption{BER curves of the system with 8 transmit and 8 receive antennas and 16QAM when divided decoding with SD is employed according to both $\{i_0=1\}$ and $\{j_0=7\}$.} \label{fig2}
\end{figure}

\begin{figure}[!t]
\centering
\includegraphics[width=2.8in, clip]{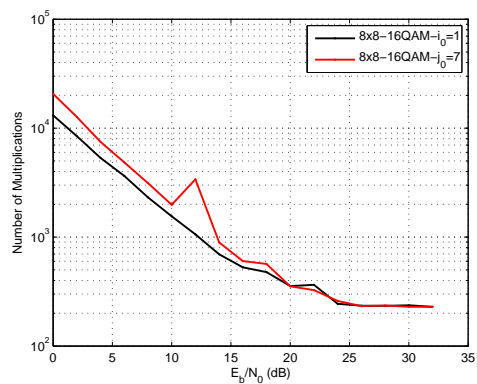}
\caption{The sample means of the number of multiplications required for divided decoding with SD for both $\{i_0=1\}$ and $\{j_0=7\}$, where the sample size is 100,000 and the system uses 8 transmit and 8 receive antennas and 16QAM.} \label{fig3}
\end{figure}

\begin{figure}[!t]
\centering
\includegraphics[width=2.8in, clip]{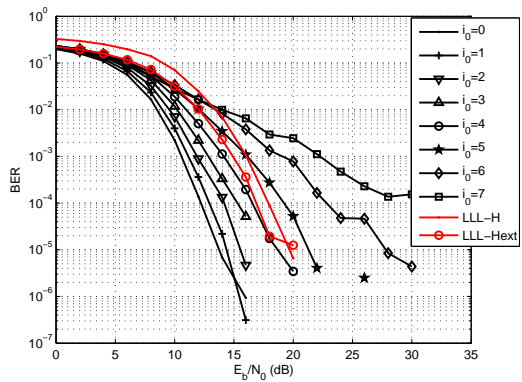}
\caption{BER curves of the system with 8 transmit and 8 receive antennas and 16QAM when divided decoding with SD is employed according to $\{i_0=0\}, \{i_0=1\}, \{i_0=2\}, \{i_0=3\}, \{i_0=4\}, \{i_0=5\}, \{i_0=6\}$, and $\{i_0=7\}$. In addition, for more effective comparison the BER curves of SIC with LLL LR applied to $\mbf{H}$, marked by `LLL-H', and SIC with LLL LR applied to $\bar{\mbf{H}}$, marked by `LLL-Hext', are included.} \label{fig4}
\end{figure}

\begin{figure}[!t]
\centering
\includegraphics[width=2.8in, clip]{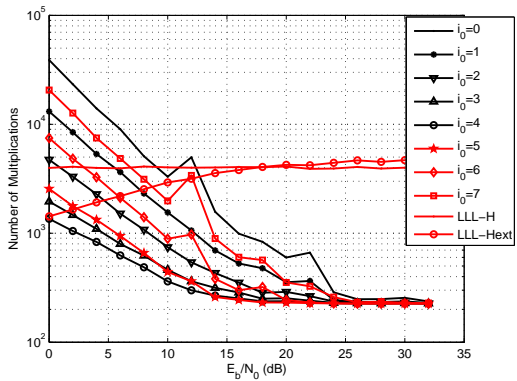}
\caption{The sample means of the number of multiplications required for divided decoding with SD according to $\{i_0=0\}, \{i_0=1\}, \{i_0=2\}, \{i_0=3\}, \{i_0=4\}, \{i_0=5\}, \{i_0=6\}, \{i_0=7\}$ and those for both SIC with LLL LR applied to $\mbf{H}$ and SIC with LLL LR applied to $\bar{\mbf{H}}$, where the sample size is 100,000 and the system uses 8 transmit and 8 receive antennas and 16QAM.} \label{fig5}
\end{figure}

We generate $\mbf{H}$ so that the entries of it have i.i.d. circularly symmetric complex normal distributions with mean zero and variance 1.0. The number of new generations of $\mbf{H}$ is 1000 and each generated $\mbf{H}$ remains fixed during 100 symbol times. The transmit data is spatially multiplexed with $n \ (n=8)$ streams and the modulation employed is 16QAM. The entries of $\mbf{n}$ are generated to be i.i.d. circularly symmetric complex normal distributions with mean zero and variance $\sigma^2$, where $\sigma^2=\frac{m}{2\cdot SNR\cdot\log_216}$, $m=8,\ SNR=E_b/N_0$. $E_b$ denotes the average energy per bit arriving at the receiver. We compare the BER curves of some typical cases and their complexities at once. The algorithm finding each sub-vector is SD and the enumeration method used in SD is the Schnorr-Euchner's.  The complexity is computed by the number of multiplications required to find solution except QRD. \figurename \ref{fig2} and \figurename \ref{fig3} show the BER and complexity curves versus SNR of Example \ref{ex1}. Obviously $P_{err}(\{1\})\ll P_{err}(\{7\})$. $G^{'}_2(7){\begin{pmatrix}\frac{\rho^2}{4\sigma^2}\end{pmatrix}}^{-1}$ is the dominant term in $P_{err}(\{7\})$ and the slope of $\log(P_{err}(\{1\}))$ is much larger than that of $\log(P_{err}(\{7\}))$. The complexity difference between the two cases is small but as predicted in Conjecture \ref{con1} the complexity of $\{i_0=1\}$ case is slightly less than that of $\{j_0=7\}$ case. As noted in the previous section, the complexities of $\{i_0=1\}$ case and $\{j_0=7\}$ case take the form of $f(1,m=8,\sigma^2)+f(n-1=7, m-1=7, \sigma^2)+4\times7$ and $f(7,8,\sigma^2)+f(1, m-n+1=1, \sigma^2)+4\times7$ respectively. And the complexity is shown to be more dependent on the first factor, the sizes of sub-vectors, than the second factor, the number of rows of the sub-matrix of $\mbf{H}$ corresponding to each sub-vector; though the effect of the second factor on the slope of BER curve is equivalent to the first factor's. Notice that the second factors of $f$ of $\{i_0=1\}$ and $\{j_0=7\}$ cases are $\{8, 7\}$ and $\{8, 1\}$ respectively and that the first factors are equal to $\{1, 7\}$. Similar phenomena appear for the pairs $(\{i_0=2\},\{i_0=6\}), (\{i_0=3\},\{i_0=5\})$ in \figurename \ref{fig4} and \figurename \ref{fig5}. On the other hand, the gap of the BER's and the slopes of BER curves between the two components composing pairs $(\{i_0=1\},\{i_0=7\}), (\{i_0=2\},\{i_0=6\}), (\{i_0=3\},\{i_0=5\})$ decreases as the index difference between the two decreases, where the index difference is equal to the difference of the two sub-vector sizes related to the pair of indices. As for complexity, Conjecture \ref{con1} is valid for limited ranges of SNR and the all curves almost coincide at high SNR range. \figurename \ref{fig5} shows also that the complexity decreases as the difference of the two sub-vector sizes related to $\{i_0\}$ decreases. But BER increases for fixed SNR and the slope of BER curve decreases, as $i_0$ increases.

\begin{figure}[!t]
\centering
\includegraphics[width=2.8in, clip]{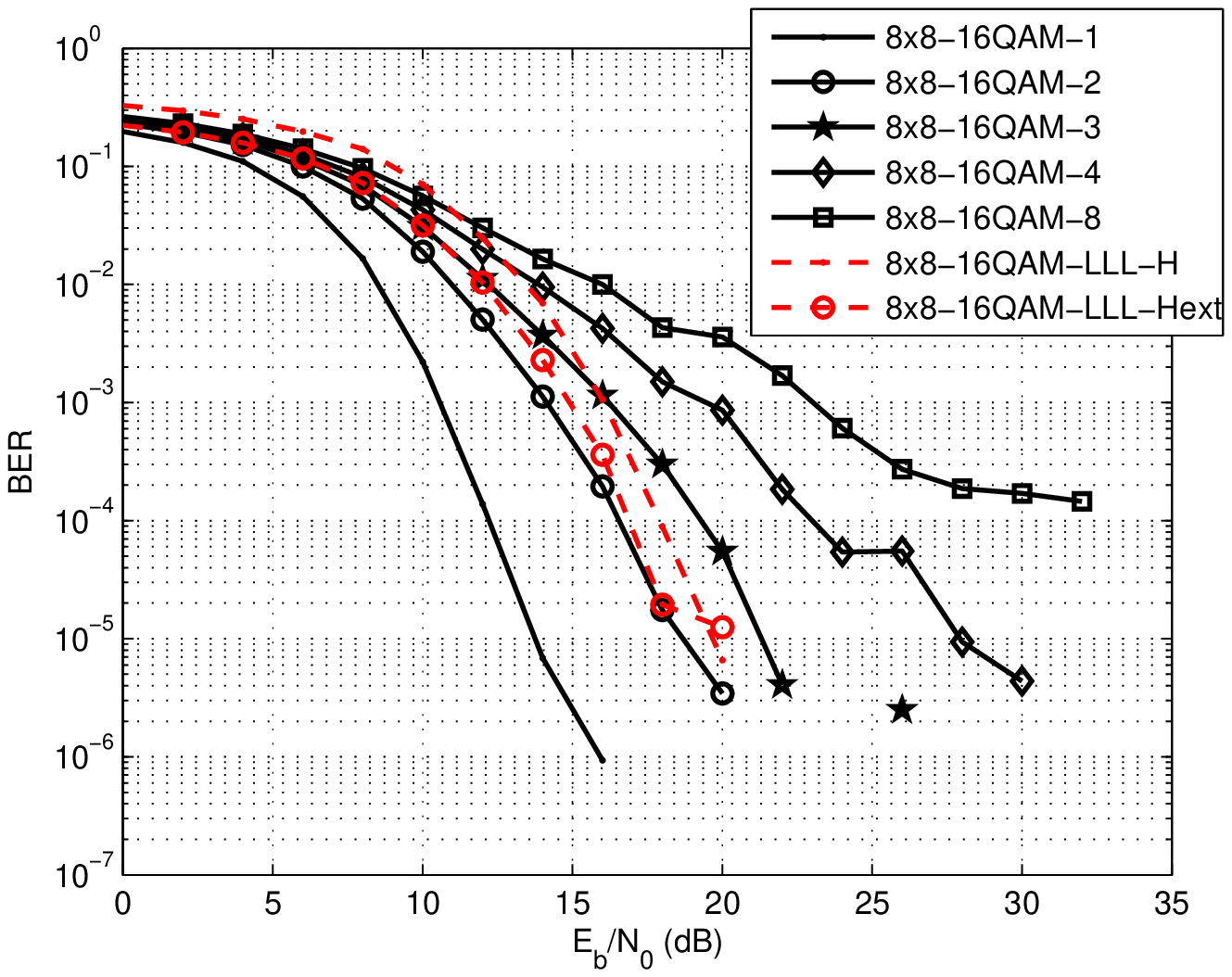}
\caption{BER curves of the system with 8 transmit and 8 receive antennas and 16QAM when DSD with $k$ split sub-systems of equal size (except for $k=3$) and $k=1,2,3,4,8$ are performed on (\ref{e3}). When $k=3$ the sub-vector sizes are $2.5,2.5,3$. For more effective comparison the BER curves of both SIC with LLL LR applied to $\mbf{H}$ and SIC with LLL LR applied to $\bar{\mbf{H}}$ are included.} \label{fig6}
\end{figure}

 In \figurename \ref{fig4} and \figurename \ref{fig5}, we compare also divided decodings according to $\{i_0\}$'s, $i_0=0,1,\dots, 7$ based on SD with both LLL LR aided ZF-SIC and LLL LR aided SIC applied to the MMSE extended system. Divided decoding according to $\{i_0=0\}$ means that the original equation is not split. Although LR aided MMSE SIC has its BER curve very close to that of SD for a system with 4-QAM modulation, 4 transmit and 4 receive antennas and still has a close BER curve to that of SD for a system with 4-QAM, 6 transmit and 6 receive antennas \cite{Wu2}, the gap gets bigger as the number of transmit (receive) antennas changes from 4 to 6. And in our simulation result with 16QAM, 8 transmit and 8 receive antennas the gap becomes more bigger, though the slopes of the BER curves of LR aided SIC's are almost the same to that of SD at a mid to high SNR range. Divided decodings according to $\{i_0\}$ based on SD for $i_0=0, 1, 2, 3, 4$ have better performances than those of LLL LR aided SIC's, and at the same time have lower computation amounts than LLL LR aided SIC's for SNR's greater than or equal to $14, 8, 5, 2, 0$ dB respectively. The number of multiplications is counted during lattice reduction process, slicing and substitutions for LLL LR aided SIC's, and for fair comparison the multiplications required for the first QRD is not counted. For SNR's greater than 12 dB, when the channel is steady for more than 10 symbol times then LLL LR aided SIC's are expected more efficient than divided decoding according to $\{i_0=4\}$ based on SD because the error rate difference is slight and lattice reduction process is not necessary for at least 10 symbol times. For SNR's less than or equal to 12dB, LR aided SIC's does not improve error rate, compared with SIC's\footnote{This claim can be verified in \figurename \ref{fig6} and \figurename \ref{fig8} for ZF-SIC and MMSE-SIC respectively. In these two figures the graphs for $k=8$ are the same with  ZF-SIC and MMSE-SIC respectively.}. If channel varies fast, divided decoding according to $\{i_0=4\}$ with SD outperforms LLL aided SIC's in both error rate and complexity. Divided decodings according to $i_0=1, 2, 3$ are also outperforming LLL aided SIC's for wide ranges of SNR.

\begin{figure}[!t]
\centering
\includegraphics[width=2.8in, clip]{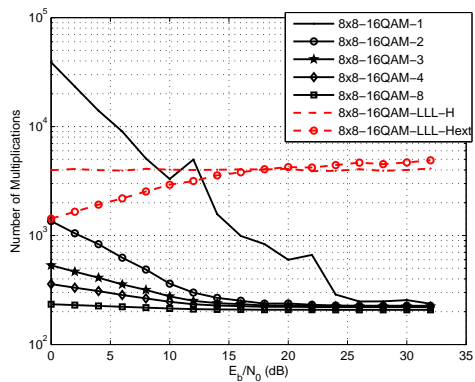}
\caption{The sample means of the number of multiplications required for DSD with $k$ split sub-systems of equal size for $k=1,2,4,8$ to find an approximate solution of (\ref{e3}) and those required for both SIC with LLL LR applied to $\mbf{H}$ and SIC with LLL LR applied to $\bar{\mbf{H}}$. When $k=3$ the sub-vector sizes are $2.5,2.5,3$.} \label{fig7}
\end{figure}

\begin{figure}[!t]
\centering
\includegraphics[width=2.8in, clip]{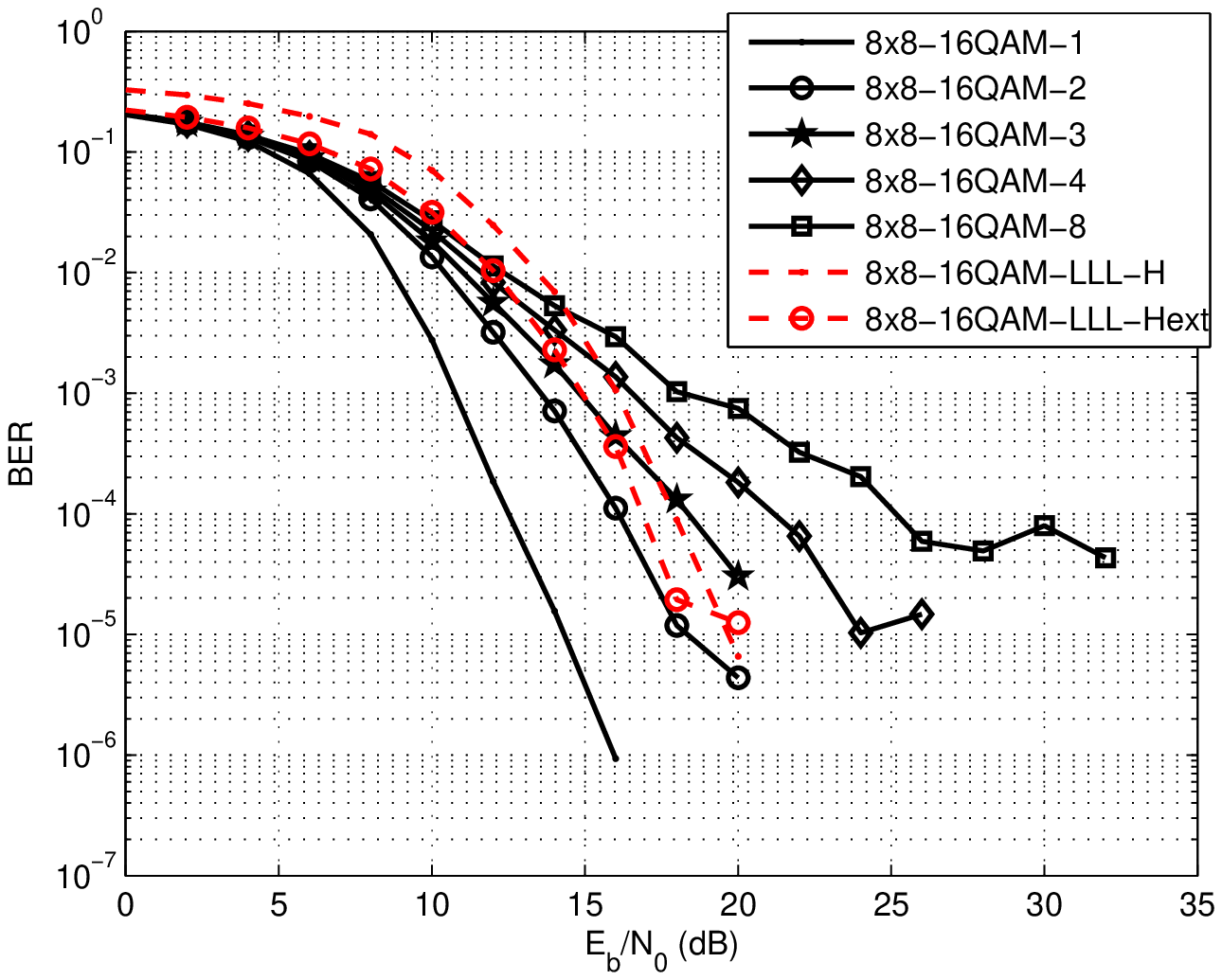}
\caption{BER curves of the system with 8 transmit and 8 receive antennas and 16QAM when DSD with $k$ split sub-systems of equal size (except for $k=3$) and $k=1,2,3,4,8$ are performed on (\ref{e9}). When $k=3$ the sub-vector sizes are $2.5,2.5,3$. For more effective comparison the BER curves of both SIC with LLL LR applied to $\mbf{H}$ and SIC with LLL LR applied to $\bar{\mbf{H}}$ are included.} \label{fig8}
\end{figure}

In \figurename \ref{fig6} and \figurename \ref{fig7} we present the BER and complexity versus SNR of divided decoding based on SD (DSD) applied to (\ref{e3}) with $k$ split of equal size, $k=1,2,3,4,8$. LLL LR aided SIC's are also compared. DSD with $k=1$ is equal to the full SD, and DSD with $k=8$ is equal to ZF-SIC. When $k=3$, the sub-vector sizes are 2.5, 2.5, and 3 where 2.5 means two complex symbols and real (or imaginary) part of a symbol. We can see the transition of BER and complexity from SD to SIC as $k$ varies from 1 to 8. For $k\geq2$, DSD with $k$ split looks better in complexity than LLL LR aided SIC's. The error rates of DSD with $k=2,3$ are near those of LLL LR aided SIC's. The decrease in complexity shrinks as $k$ increases because the decrease in sub-vector size diminishes.

\begin{figure}[!t]
\centering
\includegraphics[width=2.8in, clip]{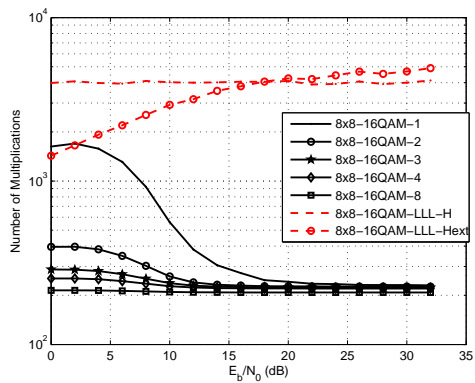}
\caption{The sample means of the number of multiplications required for divided decodings based on SD with $k$ split sub-systems of equal size (except for $k=3$) for $k=1,2,3,4,8$ to find an approximate solution to $\hat{\mbf{s}}_M$ and those required for both SIC with LLL LR applied to $\mbf{H}$ and SIC with LLL LR applied to $\bar{\mbf{H}}$. When $k=3$ the sub-vector sizes are $2.5,2.5,3$. } \label{fig9}
\end{figure}

BER curves of DSD applied to (\ref{e9}), which we call DMSD, in \figurename \ref{fig8} show that DMSD has better performance than DSD for each $k$ except $k=1$, when $k=1$ the error rates are almost the same. The trends of BER increase and complexity decrease for DMSD appeared in \figurename \ref{fig8} and \figurename \ref{fig9} respectively are similar to DSD, but the increase and decrease rates are smaller than those of DSD. Considering both BER and complexity, DMSD with proper choice of $k$ according to SNR range is expected to be better than LLL LR aided SIC's. Even DMSD with $k=1$ has lower complexity than LLL LR aided SIC's for $E_b/N_0 \ge 12dB$ even when the channel is block fading and steady for 10 symbol times. DMSD with $k=1$ is better in complexity than LLL LR aided SIC's for $E_b/N_0> 2dB$ when the channel is fast fading. DMSD with $k \ge 2$ is obviously better in complexity than LLL LR aided SIC's, and DMSD with $k \le 3$ is no worse than LLL LR aided SIC's in error rate.

The BER curves in \figurename \ref{fig2}, \figurename \ref{fig4}, \figurename \ref{fig6}, and \figurename \ref{fig8} present the diversity order transitions which are analyzed in the former section. The transitions of complexity curves in \figurename \ref{fig3}, \figurename \ref{fig5}, \figurename \ref{fig7}, and \figurename \ref{fig9} correspond to Theorem \ref{thmtttt}, to some degree.

\section{Conclusion} \label{sec6}

Divided decoding offers diverse pairs of error rate and complexity for a given mother algorithm which has ML performance or near ML performance. Upper bounds of error rates and diversity orders of DSD for typical system models are obtained, from which we are assured that in many cases splitting the equation in consideration according to $n_1\leq n_2\leq\cdots\leq n_{k+2}$ is a best strategy when divided decoding with fixed sub-vector sizes $\{n_f\}_{f=1}^{k+2}$ is applied. Divided decoding controls the exponent, the number of added terms, or the bases appeared in the calculation of complexity and shows the trade-off between error rate and complexity. On the basis of this observation, we can design advanced decoding algorithms flexible in complexity and error rate by using divided decoding. We observe that DMSD is better than DSD in both error rate and complexity if we know SNR. In comparison with LLL LR aided SIC's, DMSD and DSD are outperforming in error rate and complexity if the channel varies fast, and still outperforming for wide ranges of SNR when the channel changes slow. For further studies, adaptive applications of divided decoding to given conditions need to be considered.


%



\ifCLASSOPTIONcaptionsoff
  \newpage
\fi



%

\newpage

\end{document}